\documentstyle[12pt]{article}
\begin{document} 
\global\parskip 6pt
\newcommand{\be}{\begin{equation}}
\newcommand{\ee}{\end{equation}}
\newcommand{\bea}{\begin{eqnarray}}
\newcommand{\eea}{\end{eqnarray}}
\newcommand{\non}{\nonumber}

\begin{titlepage}
\vspace*{1cm}
\begin{center}
{\Large\bf Entropy of Three-Dimensional Black Holes\\
in String Theory}\\
\vspace*{2cm}
Danny Birmingham\footnote{Email: dannyb@ollamh.ucd.ie}\\
{\em Department of Mathematical Physics,\\
University College Dublin,\\
Belfield, Dublin 4, Ireland}\\
\vspace{.5cm}
Ivo Sachs\footnote{Email: ivo.sachs@durham.ac.uk}\\
{\em Department of Mathematical Sciences,\\
University of Durham,\\
South Road, Durham City, DH1 3LE, United Kingdom}\\
\vspace*{.5cm}
Siddhartha Sen\footnote{Email: sen@maths.tcd.ie}\\
{\em Department of Mathematics,\\
Trinity College Dublin,\\
Dublin 2, Ireland}\\
\vspace{1cm}
\begin{abstract}
It is observed that the three-dimensional BTZ black hole is a
supersymmetric solution of the low-energy field equations of
heterotic string theory compactified on an Einstein space.
The solution involves a non-zero dilaton and NS-NS H-field.
The entropy of the extreme black hole can then be computed
using string theory and the asymptotic
properties of anti-de Sitter space,
without recourse to a D-brane analysis.
This provides an explicit example of a black hole
whose entropy can be computed using fundamental string theory,
as advocated by Susskind.
\end{abstract}
\vspace{1cm}
January 1998
\end{center}
\end{titlepage}

\section{Introduction}
Recently, there has been considerable interest in the
microscopic derivation
of the Bekenstein-Hawking entropy formula for black holes
in string theory \cite{SV}-\cite{Maldacena1}.
As pointed out in \cite{Horowitz}, for example, a key property
in trying to compute the entropy of a string theory black hole
is the presence of supersymmetry. However, as observed there,
the known supersymmetric black holes in string theory typically
have zero horizon area at extremality, unless a number
of RR fields are excited. With the understanding
that D-branes are the carriers of RR charge \cite{Polchinski},
it has become possible to
reliably compute the entropy in certain examples.
However, according to the philosophy of \cite{Susskind}, it
would be appealing to see directly an example
of a supersymmetric string theory black hole which
has a non-zero horizon area at extremality,
without the presence of RR fields,
and with the entropy computable directly
in terms of fundamental string states.
The purpose of the present note is to observe that indeed
there does exist a three-dimensional black hole with these properties.
The black hole under consideration is the Ba\~{n}ados-Teitelboim-Zanelli
(BTZ) black hole \cite{BTZ,BHTZ},
which has the local geometry of anti-de Sitter
spacetime, $\mathrm{adS_{3}}$; for a review, see \cite{Carlip}.
A further exploration of the nature of black holes within
the context of conformal field theory was presented in
\cite{Strominger1}, where the decay rate of certain four- and
five-dimensional black holes was understood from the perspective
of conformal field theory. This viewpoint was analyzed
for the BTZ black hole in \cite{BSS}.

In order to understand the entropy of the BTZ black hole from
the point of view of string theory, we must first show
that it can be obtained as a compactified solution
of string theory. In \cite{Kaloper,HorRevLett,kumar},
the BTZ black hole was given an interpretation as a solution
of the three-dimensional string action with non-zero
cosmological constant.
Our aim here is different, however; we seek
a compactified solution of the low-energy $10$-dimensional string
equations of motion, of the form $\mathrm{adS_{3}}\times K_{7}$,
where $K_{7}$ is a compact internal space.
In order to achieve this, one requires a non-constant dilaton,
in contrast to the solution presented in \cite{Kaloper,HorRevLett,kumar}.
We should remark here that the solution presented is precisely
that given in \cite{Duff1}; see \cite{Duff2} for a review.

The main observation here is to see that by compactifying
heterotic string theory on a round $S^{7}$, for example,
which has trivial holonomy group,
we obtain an $N=8$ theory defined on $\mathrm{adS_{3}}$. The BTZ
black hole is then given by the standard identification
of $\mathrm{adS_{3}}$ \cite{BHTZ,Carlip}.
The only non-zero fields are the metric, the dilaton,
and the NS-NS $3$-form $H$.
We thus obtain a concrete example of a black hole which has
non-zero horizon length at extremality, and which can be
obtained as a compactified supersymmetric
solution of string theory without RR fields.
We will confine our attention to the case of the extreme black hole,
and thus we seek a compactification which preserves supersymmetry, since
it has been shown in \cite{CH} that the BTZ black hole is supersymmetric
if and only if it is extreme.
Furthermore, it was observed in \cite{Brown} that the asymptotic symmetry
algebra of three-dimensional anti-de Sitter gravity
consists of left-moving and right-moving Virasoro algebras,
with the mass of the black hole being related to
the Hamiltonian.
Thus, based on string theory and supersymmetry,
and the special asymptotic properties
of three-dimensional anti-de Sitter space,
we have a realization of the  argument presented in \cite{Susskind}.

\section{Compactification of String Theory on $\mathbf{adS_{3}}$}
In this section, we review the solution presented in \cite{Duff1}.
We seek a compactification of string theory to a product of
three-dimensional anti-de Sitter spacetime with a compact internal
space. We write the the
$10$-dimensional coordinates as $x^{M} = (x^{\mu},y^{m})$,
with $\mu=0,1,2$, and $m=3,...,9$.
We follow the conventions of \cite{Duff2}, and use the gamma
matrix representation presented in \cite{Bakas}.
The action of $N=1$ $d=10$ supergravity in the Einstein frame
is given by
\bea
S(\mathrm{Einstein\; frame}) =
\frac{1}{2\kappa^{2}}\int\;d^{10}x\;\sqrt{-\hat{g}}\left(
\hat{R} - \frac{1}{2}(\hat{\nabla}\phi)^{2} - \frac{1}{12}e^{-\phi}
\hat{H}^{2}\right).
\eea
Here, $\phi$ is the dilaton field, and the $3$-form field
$H_{MNP}$ is the NS-NS field which couples to the string.
It is convenient in the following to use the so-called
$5$-brane frame, with metric $g_{MN}$ conformally related
to the Einstein metric $\hat{g}_{MN}$ by
\be
\hat{g}_{MN} = e^{\phi/6}g_{MN}.
\ee
The action in the $5$-brane frame then takes the form
\bea
S(5-\mathrm{brane\; frame}) = \frac{1}{2\kappa^{2}}
\int\;d^{10}x\;\sqrt{-g}\left(
\Phi R - \frac{1}{12}\Phi^{-1}H^{2}\right),
\eea
where $\Phi = e^{2\phi/3}$.
The equations of motion in the $5$-brane frame are given by
\bea
R_{MN} = \Phi^{-1}\left(\nabla_{M}\nabla_{N}\Phi
- g_{MN}\nabla^{2}\Phi\right)
&+&\frac{1}{4}\Phi^{-2}\left(H_{MPQ}H_{N}^{\phantom{N}PQ}
-\frac{1}{3}g_{MN}H^{2}\right),
\label{eq1}
\eea
\bea
\nabla_{M}\left(\Phi^{-1}H^{MNP}\right) =0,
\label{eq2}
\eea
\bea
R= -\frac{1}{12}\Phi^{-2}H^{2}.
\label{eq3}
\eea
We wish to obtain a product metric of the form
\bea
g_{\mu\nu} = g_{\mu\nu}(x),\;\;g_{mn}=g_{mn}(y),\;\;g_{\mu m}=0.
\eea
To obtain the required solution, we let the dilaton depend
only on the spacetime coordinates, i.e., $\Phi = \Phi(x)$,
and we make the ansatz
\bea
H_{\mu\nu\rho} = A\Phi\epsilon_{\mu\nu\rho},
\label{ansatz}
\eea
where $A$ is a constant, and all other components are zero.
Clearly, (\ref{eq2}) is automatically satisfied by this
ansatz.
Taking the trace of (\ref{eq1}), and using (\ref{eq3}), we find
\bea
\nabla^{2}\Phi = -\frac{1}{18}\Phi^{-1}H^{2}.
\eea
Hence, using (\ref{ansatz}), we find
\bea
\left(\nabla^{2} - \frac{A^{2}}{3}\right) \Phi = 0.
\label{phi}
\eea
To solve the remaining equation,
we impose the condition \cite{Duff1,Duff2}
\bea
\nabla_{\mu}\nabla_{\nu}\Phi = \frac{1}{3}g_{\mu\nu}\nabla^{2}\Phi.
\eea
Then, (\ref{eq1}) yields
\bea
R_{\mu\nu} &=& -\frac{2A^{2}}{9}g_{\mu\nu},\label{ads}
\eea
\bea
R_{mn} &=& \frac{A^{2}}{6}g_{mn},
\label{int}
\eea
with $R_{\mu m}=0$.
Choosing $A=3/\ell$ yields the normalization of \cite{Carlip}.
Thus, we have obtained a solution of string theory
compactified to a product of three-dimensional anti-de Sitter spacetime
with a compact seven-dimensional Einstein space.

In order to show that the solution is a supersymmetric solution,
we must show that the Killing spinor equations are satisfied.
Namely, we must show that
the supersymmetry variations of the fermionic fields
vanish in the compactified background.
In the Einstein frame, the supersymmetry transformations are \cite{Duff2}
\bea
\delta\hat{\psi}_{M} = \hat{\nabla}_{M}\hat{\epsilon}
+ \frac{1}{96}e^{-\phi/2}\left(
\hat{\Gamma}_{M}^{{\phantom M}NPQ}
- 9\delta_{M}^{\phantom{M}N}\hat{\Gamma}^{PQ}\right)
H_{NPQ}\hat{\epsilon},
\eea
\bea
\delta \hat{\lambda} = -\frac{1}{2\sqrt{2}}
\left(\hat{\Gamma}^{M}\hat{\nabla}_{M}\phi\right)\hat{\epsilon}
+\frac{1}{24\sqrt{2}}e^{-\phi/2}\hat{\Gamma}^{MNP}H_{MNP}\hat{\epsilon}.
\eea
To transform these to the $5$-brane frame, we note that
with $\hat{g}_{MN} = e^{\phi/6}g_{MN}$, we have
\bea
\hat{\nabla}_{M} = \nabla_{M} + \frac{1}{24}\Gamma_{M}^{\phantom{M}N}
\nabla_{N}\phi.
\eea
We define
\bea
\epsilon &=& e^{-\phi/24}\hat{\epsilon},\non\\
\psi_{M} &=& e^{-\phi/24}\left(\hat{\psi}_{M} +
\frac{1}{6\sqrt{2}}\hat{\Gamma}_{M}\hat{\lambda}\right),\non\\
\lambda &=& e^{\phi/24}\hat{\lambda}.
\eea
Then, in the $5$-brane frame, the supersymmetry transformations
take the form
\bea
\delta{\psi}_{M} = \nabla_{M}\epsilon
+ \frac{1}{96}e^{-2\phi/3}\left(
\Gamma_{M}^{{\phantom M}NPQ} - 9\delta_{M}^{\phantom{M}N}\Gamma^{PQ}
+\frac{1}{3}\Gamma_{M}\Gamma^{NPQ}\right)
H_{NPQ}\epsilon,
\eea
\bea
\delta \lambda = -\frac{1}{2\sqrt{2}}
\left(\Gamma^{M}\nabla_{M}\phi\right)\epsilon
+\frac{1}{24\sqrt{2}}e^{-2\phi/3}\Gamma^{MNP}H_{MNP}\epsilon.
\eea
A  representation of the gamma matrices relevant to the
$3+7$ split is given in \cite{Bakas}.
We write the $10$-dimensional spinor
as $\epsilon(x,y) = \eta(x)\otimes\chi(y)$.
Then, we find
\bea
\delta \psi_{\mu} = \nabla_{\mu}\eta -\frac{A}{6}J\gamma_{\mu}\eta = 0,
\label{susy1}
\eea
\bea
\delta \psi_{m} = \nabla_{m}\chi \pm i\frac{A}{12}\Sigma_{m}\chi = 0,
\label{susy2}
\eea
\bea
\delta \lambda = -\frac{1}{2\sqrt{2}}\left(\gamma^{\mu}
\nabla_{\mu}\phi\right)\eta
+ \frac{A}{4\sqrt{2}}J\eta = 0,
\label{susy3}
\eea
with $\gamma^{4}J\eta = \pm i\eta$, where $\gamma^{4}$ and $J$ are
defined in \cite{Bakas}. We also note that $\gamma_{\mu}$ and
$\Sigma_{m}$ are the three- and seven-dimensional gamma matrices,
respectively.
The integrability conditions implied by
(\ref{susy1}) and (\ref{susy2}) are precisely the conditions
(\ref{ads}) and (\ref{int}), respectively.
Also, eqns. (\ref{susy1}) and (\ref{susy3}) imply that
\bea
(\nabla\phi)^{2} = \frac{A^{2}}{4},\;\;\nabla^{2}\phi = \frac{A^{2}}{3},
\eea
which in turn is precisely the condition (\ref{phi}).
Hence, we see that the above solution is indeed a supersymmetric
solution of low-energy string theory.
Indeed, this is precisely the solution presented in
\cite{Duff1,Duff2}, given there in terms of the dual $7$-form field
defined by $H = e^{\phi}{^{*}K}$.
The internal space must be an Einstein space, and
the number of supersymmetries in three dimensions depends
on the holonomy of the internal space. For the case
of seven-manifolds, this has been analyzed in some detail
within the context of compactifications of $11$-dimensional
supergravity \cite{Duff3}.
We note here that by choosing the round metric on $S^{7}$
which has trivial holonomy, we obtain
an $N=8$ supersymmetric theory in three dimensions.

\section{The Entropy}
In units with $8G=1$, the entropy of the BTZ black hole
is given by \cite{Carlip}
\bea
S = 4\pi r_{+},
\eea
where
\bea
r_{+}^{2} = \frac{M\ell^{2}}{2}\left\{1\pm\left[1
-\left(\frac{J}{M\ell}\right)^{2}\right]^{1/2}\right\}.
\eea
For the extreme (BPS) black hole, we have $J=\pm M\ell$.
Hence, the entropy is given by
\bea
S =  2\pi\sqrt{2M\ell^{2}}.
\eea
Now, according to \cite{Brown,CH}, we have
\bea
M\ell= \overline{L}_{0} + L_{0},\;\; J = \overline{L}_{0} - L_{0},
\eea
where the left and right Virasoro generators $L_{0}$
and $\overline{L}_{0}$ have central charge
$c=12\ell$.
Such a central charge can be realized by an $N=8$ superconformal
field theory, and thus we see that the $S^{7}$ compactification is one way
to achieve this.
Hence, in the extreme case $J= M\ell$, we have  $L_{0}=0$,
and $\overline{L}_{0} = M\ell$.
Also, in the extreme case $J = -M\ell$, we have $\overline{L}_{0} =0$,
and $L_{0} = M\ell$.
For very massive black holes, the degeneracy  of left-moving
and right-moving string
states is given by \cite{Cardy}
\bea
S = 2\pi\sqrt{\frac{cn_{L}}{6}} = 2\pi\sqrt{\frac{cn_{R}}{6}}
= 2\pi\sqrt{2M\ell^{2}},
\eea
where $n_{L}(n_{R})$ is the eigenvalue of $L_{0}(\overline{L}_{0})$.
We thus have agreement between the string entropy and the
Bekenstein-Hawking area formula.

\section{Conclusions}
The relevance of three-dimensional anti-de  Sitter spacetime to the
microscopic derivation of the entropy of certain higher-dimensional
black holes in string theory has been discussed in \cite{Sfetsos}.
The derivation presented in \cite{Sfetsos} relies on the
entropy calculation of the BTZ black hole given in \cite{Carlip2}.
In \cite{hyun}, the emergence of the BTZ black hole
from $D$-brane configurations of higher-dimensional black holes
was discussed.
It has also been shown in \cite{Maldacena2} that $\mathrm{adS_{3}}$
plays a role in the study of a certain limit of
superconformal field theories.
Recently, there has been a construction of four- and five-dimensional
anti-de Sitter black holes \cite{Vanzo}-\cite{Banados}.
In particular, it would be very interesting to
understand the entropy of the four-dimensional case from the point
of view of M Theory. In this regard, we observe that
there is the well-known compactification of 11-dimensional supergravity
to $\mathrm{adS_{4}}$ \cite{Duff3}.
Also, in the five-dimensional case, it has been shown that
the entropy is not proportional to the area \cite{Banados}, and it would
be interesting to
see this result emerge from string theory.

\noindent{\large\bf Note Added}\\
After this work was completed, we
noticed Ref.\cite{Strominger2}. The present
work provides a concrete realization of the observation of
Strominger.

\noindent{\large\bf Acknowledgements}\\
S. Sen would like to thank S. Das for drawing Ref.\cite{Strominger2}
to his attention.
 
\end{document}